\documentclass[11pt]{article}
\pdfoutput=1

\usepackage[final]{acl}

\usepackage{times}
\usepackage{latexsym}
\usepackage{amsmath}
\usepackage{amssymb}
\usepackage{booktabs}
\usepackage{multirow} 
\usepackage{bm}
\usepackage{enumitem}
\usepackage{hyperref}

\usepackage[T1]{fontenc}

\usepackage[utf8]{inputenc}

\usepackage{microtype}

\usepackage{inconsolata}

\usepackage{graphicx}

\title{Rhythm Controllable and Efficient Zero-Shot Voice Conversion \\ via Shortcut Flow Matching}

\author{
Jialong Zuo$^{1}$\thanks{Equal contribution.}, Shengpeng Ji$^{1}$\footnotemark[1], Minghui Fang$^{1}$\footnotemark[1], Mingze Li$^{1}$, Ziyue Jiang$^{1}$, Xize Cheng$^{1}$ \\
\textbf{Xiaoda Yang$^{1}$,Chen Feiyang$^{2}$,Xinyu Duan$^{2}$, Zhou Zhao$^{1}$\thanks{Corresponding author.}} \\
$^{1}$Zhejiang University\hspace{1em}  $^{2}$Huawei Cloud\\
\href{mailto:jialongzuo@zju.edu.cn}{\texttt{jialongzuo@zju.edu.cn}} \hspace{1em} 
\href{mailto:zhaozhou@zju.edu.cn}{\texttt{zhaozhou@zju.edu.cn}}
}

\begin{document}
\maketitle
\begin{abstract}
Zero-Shot Voice Conversion (VC) aims to transform the source speaker’s timbre into an arbitrary unseen one while retaining speech content. Most prior work focuses on preserving the source's prosody, while fine-grained timbre information may leak through prosody, and transferring target prosody to synthesized speech is rarely studied. In light of this, we propose R-VC, a rhythm-controllable and efficient zero-shot voice conversion model. R-VC employs data perturbation techniques and discretize source speech into Hubert content tokens, eliminating much content-irrelevant information. By leveraging a Mask Generative Transformer for in-context duration modeling, our model adapts the linguistic content duration to the desired target speaking style, facilitating the transfer of the target speaker's rhythm. Furthermore, R-VC introduces a powerful Diffusion Transformer (DiT) with shortcut flow matching during training, conditioning the network not only on the current noise level but also on the desired step size, enabling high timbre similarity and quality speech generation in fewer sampling steps, even in just two, thus minimizing latency. Experimental results show that R-VC achieves comparable speaker similarity to state-of-the-art VC methods with a smaller dataset, and surpasses them in terms of speech naturalness, intelligibility and style transfer performance. Audio samples are available at \url{https://r-vc929.github.io/r-vc/}.

\end{abstract}

\section{Introduction}
Voice conversion (VC) aims to transform the speaker’s timbre in speech to match that of a target speaker while preserving the original speech content. The core approach involves disentangling speech into several individual components, such as linguistic content, timbre, and prosody or style. The speaker conversion is achieved by integrating the linguistic content of the source speech with the target speaker characteristics. Some previous methods focus on disentangling speech content and speaker characteristics. Information bottleneck-based methods \cite{hsu2016voice,qian2019autovc,qian2020unsupervised} are developed to disentangle speakers and content. Some studies \cite{choi2021neural,qian2022contentvec,ning2023expressive} also use signal perturbation techniques to alter speech utterances to make it speaker irrelevant before content extraction. Vector quantization methods like K-means \cite{hsu2021hubert} or VQ-VAE \cite{van2017neural,wu2020vqvc+} have been employed to discretize content features, effectively removing speaker variability. Recently, advanced methods \cite{van2022comparison,yao2024promptvc,choi2023diff} have directly utilized features extracted from pre-trained self-supervised speech representation networks \cite{hsu2021hubert,baevski2020wav2vec,babu2021xls} as linguistic content. However, although these features provide rich semantic information, the retained residual timbre information may lead to timbre leakage.



Inspired by the powerful zero-shot capabilities of generative models \cite{ho2020denoising,peebles2023scalable,tong2023conditional,lee2023hierspeech++,jiang2023fluentspeech} and speech language models (SLMs) \cite{borsos2023audiolm,vall-e}, recent studies are proposed to enhance the naturalness and speaker similarity of the converted speech. LM-VC \cite{wang2023lm} and Uniaudio \cite{yang2023uniaudio} explore the feasibility of LMs for zero-shot voice conversion. Some diffusion-based VC \cite{popov2021diffusion,choi2024dddm} are proposed to generate natural and high-quality speech. Recent advancements in conditional flow matching methods \cite{chen2024disentangling,du2024cosyvoice,zuo2025enhancing} leverage in-context learning (ICL) capability through target speech prompting, resulting in improved speaker similarity. Nevertheless, these methods have predominantly focused on maintaining the source speech’s prosody which often leads to two issues: the coupling between timbre and prosody may result in timbre leakage, and these models struggle to transfer prosody or rhythm to align with the target speech. Moreover, although diffusion or flow matching offers advantages in generating high-quality speech, the multi-step sampling paradigm (typically $\geq$ 10 steps) results in relatively high latency, which, compared to non-diffusion methods, limits its practical application in real-world scenarios.

To address these challenges, we propose R-VC, a rhythm controllable voice conversion system that enables efficient zero-shot speaker conversion while replicating the target speech's rhythm to produce natural, high-quality speech with minimal generation steps. Specifically, to obtain speaker-irrelevant linguistic content and mitigate timbre leakage, we apply data perturbation techniques to the input waveform and utilize pretrained HuBERT and K-Means models to extract discrete content tokens. To further eliminate prosodic information, we deduplicate the content tokens and introduce a mask transformer based duration model that iteratively predicts masked durations based on contextual information, effectively learning the target speaker’s rhythm. Additionally, we capture both time-invariant and time-varying speaker characteristics through a diffusion transformer which employs a target speech prompting strategy to enhance timbre similarity. More importantly, a shortcut flow matching technique is introduced in the DiT decoder during the training phase, enabling the model to accurately skip ahead in the denoising process by conditioning on the assigned step size. The main contributions are summarized as follows:
\begin{itemize}[itemsep=1pt]
    \item We present R-VC, a robust and fast zero-shot voice conversion system that enables flexible control over rhythm and speaker identity.
    \item We introduce a masked generative transformer model for in-context duration modeling, capable of adapting the duration of the same linguistic content to different styles. Additionally, we explore the effects of different duration granularities on the performance of VC models, like token-level and sentence-level.
    \item We propose a shortcut DiT-based conditional flow matching model that enables efficient few-step and even one-step generation, significantly improving generation efficiency.
    \item R-VC achieves comparable timbre similarity to SOTA models with less training data while surpassing baseline models in terms of speech naturalness, intelligibility and emotion style transfer performance. The method accelerates speech generation by 2.83 times compared to the 10-step sampling Conditional Flow Matching (CFM) methods while maintaining similar speech quality.
\end{itemize}
\section{Related Work}
\subsection{Speaker Representation in VC}
Modeling speaker identity is crucial for voice conversion, posing challenges in learning robust speaker representations and designing efficient conditioning strategies. Traditional approaches typically rely on global speaker embeddings extracted from pre-trained speaker verification (SV) models \cite{snyder2018x,casanova2022yourtts,gu2021mediumvc,qian2020unsupervised}, speaker encoders \cite{unitspeech,qian2019autovc,yang2024mscenespeech}, or advanced speaker representation models \cite{tan2021zero,cooper2020zero}, casting them as time-invariant representations. FACodec \cite{ns3} introduces a decoupled speech codec that also facilitate VC, utilizing a timbre extractor and gradient reversal layer (GRL) to derive speaker representations. Diff-HierVC \cite{choi2023diff} introduces a diffusion based pitch generator for F0 modeling and a style encoder for global speaker representation. However, these methods often struggle in zero-shot voice conversion scenarios due to the inherent generalization limitations of global embeddings.


\begin{figure*}[ht]
\centering
\includegraphics[height=8.0cm, width=14cm]{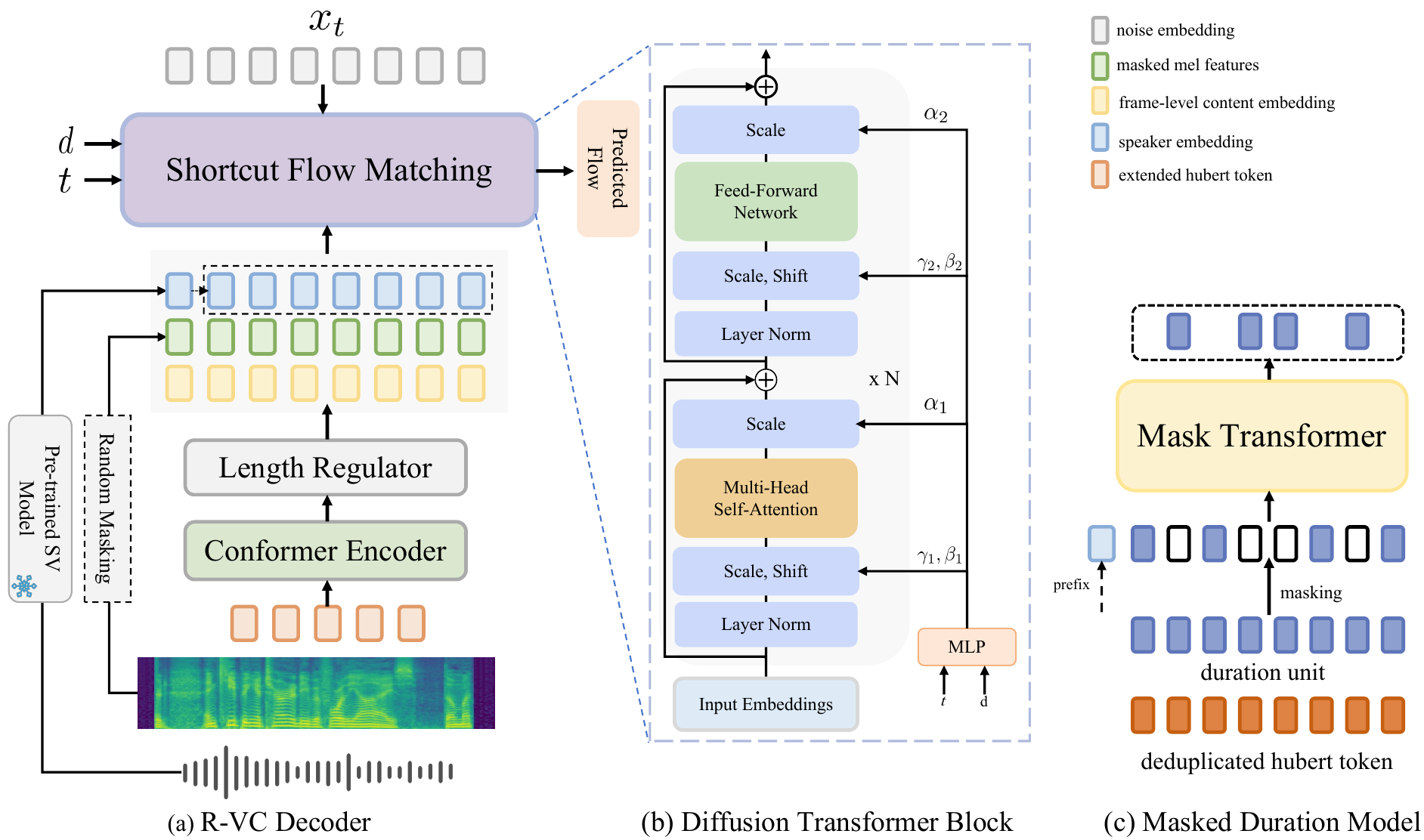}

\caption{Figure (a) illustrates the shortcut DiT-based flow matching decoder, which regresses the conditional transport vector field from noise to spectrogram while conditioning on the desired step size. Figure (b) presents the detailed structure of the diffusion transformer block. In Figure (c), the non-autoregressive Masked Duration Model is trained to predict masked duration units based on content, contextual duration, and the prefix speaker embedding.}
\label{rvs_overall}
\end{figure*}

Recently, more efficient speaker modeling techniques are also proposed. SEF-VC \cite{li2024sef} proposes a position-agnostic cross-attention mechanism to learn and incorporate speaker timbre from reference speech while RefXVC \cite{zhang2024refxvc} combines global and local embeddings to capture timbre variations. Despite these impressive results, there is still room for improvement in terms of speech naturalness and timbre similarity. Motivated by the recent success of generative models with in-context learning capabilities in speech synthesis \cite{ns3,chen2024f5,jiang2024mega} and audio generation \cite{yang2023uniaudio,borsos2023audiolm}, some studies have explored their potential in voice conversion. CosyVoice \cite{du2024cosyvoice} and ICL-VC \cite{chen2024disentangling} employ a flow-matching generative model training strategy with masking and reconstruction, utilizing the entire reference speech context to capture fine-grained speaker characteristics, significantly improving the speaker similarity. Nevertheless, these methods retain source prosody, with CosyVoice utilizing semantic tokens that contain partial style information, and ICL-VC leveraging embeddings from a pre-trained emotion model, potentially causing timbre leakage. Moreover, they fail to replicate the target speaker's rhythm. 

\subsection{Flow Matching for Speech Generation} 
Flow matching (FM) models \cite{lipman2022flow} the vector field of transport probability path from noise to data samples. Unlike diffusion-based methods like DDPM \cite{ho2020denoising}, flow matching offers more stable training and superior performance. Crucially, by leveraging ideas from optimal transport \cite{tong2023improving}, FM can be set up to yield ODEs that have simple vector fields that change little during the process of mapping samples from the source distribution onto the data distribution, since it essentially just transports probability mass along straight lines, which greatly reduces the required number of sampling steps. Recent advancements in flow matching-based generative models have demonstrated significant success not only image generation \cite{esser2024scaling} but also text-to-speech \cite{mehta2024matcha,du2024cosyvoice,le2024voicebox,jiang2025megatts}. Although FM-based methods generate high-quality speech with fewer sampling steps than traditional diffusion models, but still require multiple forward passes (e.g., 10, 25, 32), making generation slow and expensive. Inspired by shortcut models in the image generation domain \cite{frans2024one}, R-VC combines the optimal-transport flow matching loss with a self-consistency loss, which enforces the model's prediction to align with the actual data distribution at larger step sizes, boosting generation efficiency significantly.

\section{R-VC}

\subsection{Overall Architecture}
The overall architecture of R-VC is shown in Figure \ref{rvs_overall}. R-VC is a fully non-autoregressive encoder-decoder voice conversion system that employs a shortcut flow matching  DiT decoder to model in-context speaker characteristics, achieving speech generation with fewer inference steps and incorporates a masked generative transformer duration model for controllable rhythm. The whole procedure can be divided into three modules. (1) \textbf{Robust Content Representation Extraction}: discrete content tokens are first extracted using a pretrained Hubert model and K-Means clustering from perturbed waveforms and then deduplicated to remove some prosodic patterns. (2) \textbf{Mask Transformer Duration Model}: unit-level durations are obtained through a mask-predict iterative decoding conditioned on the deduplicated content, unmasked target units, and speaker embedding. (3) \textbf{Shortcut Flow Matching Estimator}: a shortcut flow matching diffusion scheme is employed to estimate the target training objectives (not only the FM objectives but also self-consistency objectives), guided by content, randomly masked context acoustic features, noise embeddings, a global speaker embedding, as well as a desired step size, propagating the model's generation capability from multi-step to few-step to one-step. During inference, the source deduplicated content tokens are extended according to the predicted durations and input into the shortcut DiT decoder to generate mel-spectrograms. Finally, a pretrained  HifiGAN \cite{kong2020hifi} vocoder synthesizes the perceptible waveform from the generated mel-spectrograms.

\subsection{Linguistic Content Representation}
In R-VC, we leverage a self-supervised learning-based pre-trained HuBERT model to extract SSL features, which are then discretized by a K-means model. The discrete tokens contain rich phonetic content information and exhibit minimal speaker variance, as demonstrated in \cite{li2024sef,soft}. Additionally, data perturbation is applied to the input waveform to eliminate content-irrelevant information. More details of the perturbation methods are in Appendix \ref{pitch_data}. 

The Hubert content tokens are extracted at a rate of 50 tokens per second (50Hz) \footnote{\url{https://github.com/facebookresearch/fairseq/blob/main/examples/hubert}}, encompassing the full duration information of the speech, similar to the SSL features from wav2vec \cite{baevski2020wav2vec} and XLS-R \cite{babu2021xls}. This means that the converted speech retains the same rhythm as the source speech, which limits the model's style transfer capability. Duration or rhythm is a key component of prosody, particularly in emotional voice conversion, where rhythm influences emotional expression \footnote{Joy is conveyed through brisk speech, while sadness is marked by a slower, more measured pace.}. Consequently, we propose to deduplicate discrete tokens to obtain unit-level durations. For example, given a speech input $S$, the content tokens from the Hubert model are represented as $C(S) = [u_1, u_1, u_1, u_2, u_3, u_3]$, which are condensed to $C'(S) = [u_1, u_2, u_3]$ with corresponding durations $D(S) = [d_1, d_2, d_3]$, where $d_1=3,d_2=1,\cdots$. This approach brings two advantages: 1) It allows for rhythm-controlled voice conversion rather than merely preserving the source speech’s rhythm; 2) It may remove style-related information, such as accent and residual variations, as discussed in \cite{lee2021textless}.

\subsection{Mask Transformer Duration Model}
Inspired by the successful application of parallel decoding in text \cite{textmask}, image \cite{maskgit}, audio \cite{soundstorm,ji2024mobilespeech} generation tasks. R-VC employs a non-autoregressive, mask-based generative transformer model for duration modeling. The non-autoregressive (NAR) duration model utilizes the mask-predict algorithm to iteratively refine unit choices, achieving high-accuracy output in just a few cycles. We sample the mask $M \in \left \{ 0,1 \right \} ^N$ according to a sine schedule for a target duration unit sequence, specifically sampling the masking ratio $p=\sin(u)$ where $u\sim \mathcal{U}\left [  0,\frac{\pi }{2}  \right ] $, the mask $M_i \sim Bernoulli(p) $ and $i\sim \mathcal{U}\left [0, N-1 \right ]$. Therefore, the masked duration sequence can be represented as $\bar{D}(S) = D(S) \odot M $, where the duration unit is preserved when $ M_i = 0 $, and replaced with a \textit{Mask} token otherwise.

As shown in Figure \ref{rvs_overall} (c). The prediction of masked target duration units is conditioned on the reduced content tokens, the unmasked portion of durations, and a global speaker embedding, where the content tokens and unmasked durations are concatenated at the dimension level, and the speaker embedding is concatenated with them at the sequence level. This prediction is modeled as:
\begin{equation}
    P(D(S) \mid \bar{D}(S), C; \theta)
\end{equation}
where $C$ denotes the conditions mentioned above. Rhythm in speech is not only content-dependent but also speaker-specific. Therefore, we introduce a deterministic speaker representation extracted from a pre-trained speaker verification model \footnote{\url{https://github.com/modelscope/3D-Speaker}} as an additional condition to achieve more accurate and efficient duration modeling. The learning objective is the cross-entropy (CE) loss between the generated and target units at masked positions:
\begin{equation}
L_{\text{CE}} = \mathbb{E} \Big( - \sum_{i=1}^{N} M_{i} \cdot \log(p(d_i \mid \bar{D}(S), C; \theta)) \Big) 
\end{equation}
We use the standard transformer block from LLAMA \cite{dubey2024llama} as the backbone for the duration model and adopt the Rotary Position Embedding (RoPE) \cite{su2024roformer} as the positional embedding, configured with fewer layers and a smaller hidden dimension. During inference, we decode the duration units in parallel through iterative decoding with a pre-defined $T$ iterations. We perform a mask operation at each iteration, followed by predict. At the initial iteration $t=0$, all the units in the sequence  $D= [d_1,\ldots,d_N]$ are masked. In subsequent iterations, $n$ units with the lowest probability scores $p$ are masked:
\begin{equation}
\bar{D}_t = \arg\min_{i}(p_i, n), \quad D_t^{'} = D \setminus \bar{D}_t,
\end{equation}
where $n$ is a function of the iteration $t$. In this work, we apply a linear decay, defined as \( n = N \cdot \frac{T - t}{T} \). After masking, the duration model predicts the target units $D_t^{'}$ based on conditions $C$ and the masked context duration units $\bar{D}_t$. For each $d_i \in D_t^{'}$, the prediction with the highest probability $p$ is selected, and the probability score is updated as follows:
\begin{align}
d_i^t &= \arg\max_w P(d_i = w \mid \bar{D}_t, C; \theta), \\
p_i^t &= \max_w P(d_i = w \mid \bar{D}_t, C; \theta).
\end{align}

This iterative approach ensures progressive refinement of the target duration sequence.

\subsection{Shortcut Flow Matching Estimator}
Drawing inspiration from previous works \cite{le2024voicebox,du2024cosyvoice}, which leveraged masking strategies for in-context learning, we explore a method that utilizes a target speech prompting strategy for zero-shot voice conversion tasks. In R-VC, an optimal-transport conditional flow matching model (OT-CFM) \cite{tong2023conditional} is employed to learn a conditional mapping from a noise distribution $x_0 \sim p_0(x)$ to a data distribution $x_1 \sim p_1(x)$. For $x$ and $t \sim U[0, 1]$, 
the OT-CFM loss function can be formulated as:
\begin{equation}
\begin{split}
L(\theta) = \mathbb{E}_{t, p_1(x_1), p_0(x_0)} \| u_t(x_t | x_1) - v_t(x_t, c; \theta) \|^2 \nonumber
\end{split}
\end{equation}
where \bm{$\theta$} represents the neural network and \bm{$c$} is the condition, composed of masked context features, content embeddings, and speaker embeddings concatenated along the channel dimension. Although the CFM model is non-autoregressive in terms of the time dimension, it requires multiple iterations to solve the Flow ODE. The number of iterations (i.e., number of function evaluations, NFE) has a great impact on inference efficiency, especially when the model scales up further. Additionally, naively taking large sampling steps leads to large discretization error and in the single-step case, to catastrophic failure. 
\begin{figure}[]
\centering
\includegraphics[height=3.0cm, width=8cm]{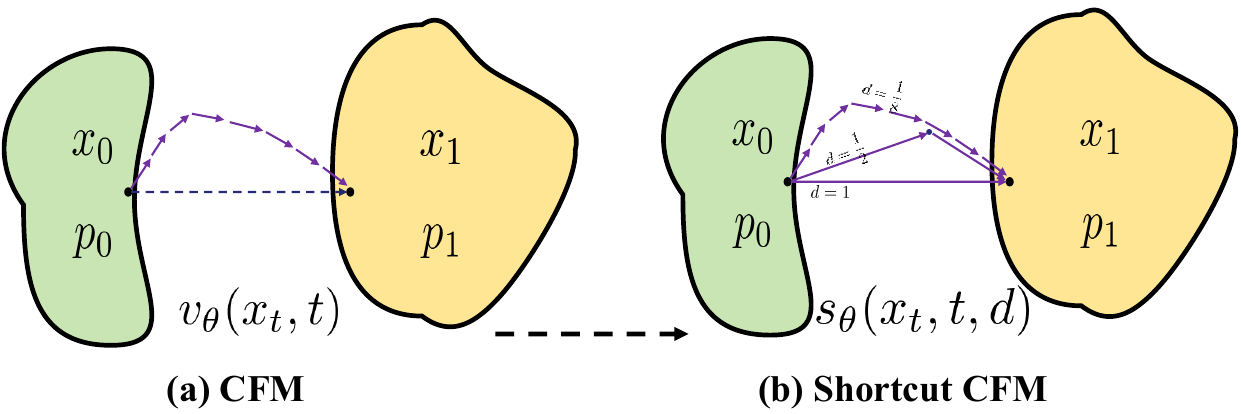}

\caption{The comparison of vanilla conditional flow matching and our shortcut flow matching methods.}
\label{shortcut_fm}
\end{figure}
Therefore, we introduce a shortcut flow matching strategy in R-VC as mentioned in \cite{frans2024one}, where an additional conditioning on the step size allows the model to account for future curvature and jump to the correct next point rather than going off track at large step sizes. Specifically, the normalized direction from $x_t$ towards the correct next point $x'_{t+d}$ is represented as the shortcut $s(x_t, t, d)$:
\begin{equation}
x^{'}_{t+d} = x_t + s(x_t, t, d) \cdot d
\end{equation}
As shown in Figure \ref{shortcut_fm}, the shortcut CFM model can be seen as a generalization of flow-matching models for larger step sizes: as $d\to0$, the shortcut becomes equivalent to the flow, while shortcut models additionally learn to make larger jumps when $d > 0$. Due to the high computational cost, it is not feasible to compute targets for training $s_\theta(x_t, t, d)$ by fully simulating the ODE forward with a small enough step size. Thanks to the consistency property of shortcut models, a self-consistency training target is proposed:
\begin{equation}
s(x_t, t, 2d) = \frac{1}{2} \left( s(x_t, t, d) + s(x'_{t+d}, t, d) \right)
\end{equation}
where one shortcut step equals two consecutive shortcut steps of half the size.
Therefore, our shortcut flow matching estimator is trained with an OT-CFM loss and a self-consistency loss as follows:
\begin{equation}
\begin{split}
L_{\small \text{S-CFM}}(\theta) = \mathbb{E}_{p_0(x_0), p_1(x_1), (t,d)} [ \| s_\theta(x_t, t, 0) -\\ (x_1 - x_0) \|^2 \, + \| s_\theta(x_t, t, 2d) - s_{\text{target}} \|^2 ] \nonumber
\end{split}
\end{equation}
Such training objective learns a mapping from noise to data which is consistent when queried under any sequence of step sizes. The more detailed information about the CFM and Shortcut CFM algorithms are in Appendix \ref{cfm_appendix}.


Regarding the backbone network for Shortcut CFM, we discarded the U-Net-style skip connection structure and opted for a Diffusion Transformer (DiT) with AdaLN-zero, unlike previous U-Net-based transformer methods \cite{mehta2024matcha,du2024cosyvoice}. The diffusion transformer is used to predict speech that matches the style of the target speaker and the content of the source discrete token. As shown in Figure \ref{rvs_overall} (a), the original HuBERT token is first encoded by a Conformer Encoder into content embeddings, which are then passed through a Length Regulator module to align with the mel-spectrogram sequence length. To enhance in-context timbre modeling, we employ a masked speech modeling approach. This includes using a random span-masked mel-spectrogram as a speaker prompt for capturing fine-grained speaker traits, along with a global speaker embedding from a pre-trained speaker verification model as a timbre conditioning. Experiments show that this method accelerates training and produces a more robust timbre representation, improving speaker similarity.
During inference, we employ the classifier-free guidance approach to steer the model $f_\theta $'s output towards the conditional generation $f_\theta(x_t, c)$ and away from the unconditional generation $f_\theta(x_t, \emptyset)$:
\begin{equation}
\hat{f}_\theta(x_t, c) = f_\theta(x_t, c) + \alpha \cdot [f_\theta(x_t, c) - f_\theta(x_t, \emptyset)]\nonumber
\end{equation}
where $c$ denotes the conditional state, $ \emptyset $ denotes the unconditional state, and $ \alpha $ is the guidance scale selected based on experimental results.
\subsection{Training and Inference}
The Mask Transformer Duration Model and the Shortcut DiT Decoder are trained independently with distinct objectives: the cross-entropy loss $L_{\text{CE}}$ and the shourtcut flow matching loss $L_{\text{S-CFM}}$, computed only at masked positions. We set $N=128$ as the smallest time unit for approximating the ODE, resulting in 8 possible shortcut lengths based on $d \in (1/128, 1/64, \dots, 1)$. Since 128 is an empirically chosen value large enough to fit the vector field of the transport probability path for $t\in[0, 1] $, smaller values lead to performance degradation, while larger values incur computational overhead. We did not explore other configurations. During each training step, we sample $x_t$, $t$, and a random $d < 1$, then perform two sequential steps with the shortcut model. The concatenation of these steps is used as the target for training at $2d$. For each training batch, a proportion $k$ of the items are used for the flow matching objective, and the remaining $(1-k)$ items for self-consistency targets.

\section{Experiments}
\subsection{Experiment Setup}
\textbf{Datasets} We utilize a subset of the English portion of the Multilingual LibriSpeech (MLS) dataset \cite{pratap2020mls}, encompassing nearly 20,000 hours and 4.8 million audio samples, to train the R-VC model and to reproduce several baseline models. For zero-shot voice conversion, we use the test-clean subset \cite{panayotov2015librispeech} across 40 speakers, while emotional voice conversion is evaluated on the ESD dataset \cite{esd}. Additionally, rhythm control is assessed with a test set of varied speaking rates from the test-clean dataset.

\textbf{Baselines}
We compare the performance of R-VC with the current state-of-the-art zero-shot VC systems, including 1) FACodec-VC \cite{ns3}; 2) CosyVoice-VC \cite{du2024cosyvoice}; 3) Diff-HierVC \cite{choi2023diff}; 4) HierSpeech++ \cite{lee2023hierspeech++}; 5) SEF-VC \cite{li2024sef}. We use the official open-source checkpoints pre-trained on large-scale datasets to generate samples for FACodec-VC, CosyVoice-VC, and HierSpeech++. For Diff-HierVC and SEF-VC, we reproduce the models using the same dataset as R-VC.
\\
\textbf{Training and Inference Setup}
We train the R-VC model on 8 NVIDIA A100 GPUs, with 600k steps for the Diffusion Transformer Decoder using a batch size of 20k Mel frames per GPU, and 100k steps for the Mask Transformer Duration Model with a batch size of 8k tokens. The Adam optimizer is employed with a learning rate of $5 \times 10^{-5} $, $ \beta_1 = 0.9 $, $ \beta_2 = 0.999 $, and 2k warmup steps. Further details are provided in Appendix \ref{details_model}.\\
\textbf{Evaluation Metrics}
We evaluate the model with objective metrics including cosine distance (SECS), character error rate (CER), word error rate (WER) for zero-shot voice conversion, and emotion score from a pre-trained emotion representation model for style transfer. Inference latency is measured by the real-time factor (RTF) on a single NVIDIA V100 GPU. For subjective evaluation, MOS assessments are conducted via Amazon Mechanical Turk, focusing on QMOS (quality, clarity, naturalness) and SMOS (speaker similarity), with automatic MOS prediction using UTMOS \cite{utmos}. Experimental details are in Appendix \ref{details_exp}.


\begin{table*}[ht]
\small
\centering
\begin{tabular}{l|ccccccc}
\toprule
\bfseries Method & \bfseries {WER$\downarrow$} &\bfseries {CER$\downarrow$} & \bfseries {SECS$\uparrow$} & \bfseries UTMOS $\uparrow$ & \bfseries QMOS $\uparrow$ & \bfseries SMOS $\uparrow$ & \bfseries RTF $\downarrow$ \\ 
\midrule
GT (Vocoder) &2.87&1.36&-&4.14&4.19$\pm$ 0.10&-&-\\
\midrule
FACodec-VC &4.68&1.96&0.902&2.88&3.21 $\pm$ 0.05&3.91 $\pm$ 0.14&\bfseries0.10\\
Diff-HierVC &4.34&2.14&0.894&3.75&3.64 $\pm$ 0.16&3.80 $\pm$ 0.06&0.47\\
HierSpeech++ &3.64&1.46&0.907&4.09&3.98 $\pm$ 0.12&3.93 $\pm$ 0.08&0.25\\
CosyVoice-VC &5.95&2.70&\bfseries0.933&4.09&3.95 $\pm$ 0.15&4.11 $\pm$ 0.16&1.74\\
SEF-VC &4.82&2.04&0.897&3.96&3.80 $\pm$ 0.07&3.85 $\pm$ 0.11&0.18\\
\midrule
R-VC (CFM, NFE=10) &\bfseries3.47&\bfseries1.39&0.931&\bfseries4.10&\bfseries4.05 $\pm$ 0.11&\bfseries4.12 $\pm$ 0.09&0.34\\
\bfseries R-VC (ours, NFE=2) &3.51&1.40&0.930&4.10&4.03 $\pm$ 0.13&4.11 $\pm$ 0.07&0.12\\
\bottomrule
\end{tabular}
\caption{The objective and subjective experimental results for zero-shot VC. \textbf{GT (Vocoder)} refers to synthesizing speech from ground truth mel features using a vocoder. \textbf{R-VC (CFM)} refers to the vanilla CFM diffusion scheme.}
\label{table1}
\vspace{-0.2in}
\end{table*}

\begin{table}[ht]
\small
\centering
\tabcolsep=2.5pt
\begin{tabular}{l|cccc}
\toprule
\bfseries Method & \bfseries {WER} & \bfseries {SECS} & \bfseries UTMOS & \bfseries{EMO} \\ 
\midrule
Target &-&-&3.93&0.692\\
\midrule
FACodec-VC &8.91&0.854&2.64&0.438\\
Diff-HierVC &10.62&0.833&3.41&0.440\\
HierSpeech++ &7.94&0.850&3.78&0.489\\
CosyVoice-VC &9.93&0.878&3.73&0.395\\
SEF-VC &9.25&0.841&3.59&0.461\\
\midrule
\bfseries R-VC (ours) &\bfseries6.95&\bfseries0.880&\bfseries3.85&\bfseries0.590\\
\bottomrule
\end{tabular}
\caption{The experimental results of emotion style transfer on an unseen ESD dataset. \textbf{EMO} refers to the average emotion score on the test set.}
\label{table2}
\vspace{-0.2in}
\end{table}

\subsection{Zero-shot VC Results}
In this section, we evaluate the zero-shot voice conversion performance of our model on an unseen test-clean dataset comprising 2620 samples. The analysis covers multiple dimensions, including speech quality, naturalness, intelligibility, timbre similarity, and inference speed. As summarized in Table \ref{table1}, our R-VC model achieves competitive timbre similarity compared to state-of-the-art methods and surpasses the baseline across all other metrics except inference latency, demonstrating its exceptional capability in zero-shot VC tasks. 

Specifically, we observe the following: 1) R-VC achieves a WER of 3.51 and a CER of 1.40, outperforming other VC systems, with the CER of the generated speech approaching GT levels, highlighting its clarity and intelligibility. 2) In terms of timbre similarity, R-VC achieves comparable performance to the CosyVoice-VC baseline, outperforming other models by a margin of 3\%. Considering the disparity in training data, where CosyVoice-VC leverages a large-scale in-the-wild dataset of 171k hours, while our model is trained on only 20k hours of MLS English data, this result highlights the effectiveness of our approach. We attribute this primarily to the powerful shortcut DiT decoder, which captures speaker characteristics by leveraging the global speaker condition and target prompt, along with the combined impact of data perturbation techniques and K-means discretization, which efficiently filters content-irrelevant information. Additionally, the superior SMOS scores confirm that our model generates speech with the highest perceived speaker similarity. 3) R-VC notably exceeds other systems in QMOS evaluations, particularly in UTMOS scores, approaching ground truth audio quality, which demonstrates its excellence in generating high-quality, clear speech. 4) Considering inference speed, by incorporating shortcut flow matching, our model achieves performance comparable to vanilla CFM while reducing sampling steps to 1/5, resulting in a 2.83× faster inference, approaching the SOTA speed of non-diffusion methods like FACodec-VC. This demonstrates the model's superior generation efficiency, making it feasible for practical applications.


\subsection{Emotion Style Transfer Ability Evaluation}
We randomly select 200 emotional speech samples from the ESD dataset to evaluate emotion style transfer. We utilize a pre-trained Emotion2Vec \cite{ma2023emotion2vec} model to calculate the emotion score for each synthesized sample, and average these scores to obtain the overall score.


As shown in Table \ref{table2}, almost all metrics for both the baseline and R-VC models show significant reductions compared to zero-shot VC results, primarily due to the lower quality of the ESD dataset and the inherent challenges of emotion style transfer. R-VC outperforms the baseline across all metrics, particularly in WER and emotion scores, demonstrating its exceptional robustness and style transfer capability. These improvements stem from two main factors: First, we have achieved cleaner linguistic content extraction, where K-means discretization and data perturbation techniques effectively eliminate irrelevant content such as timbre and style. In contrast, baseline models such as FACodec-VC rely on prosody codes from source speech, retaining source style rather than transferring it. Additionally, CosyVoice-VC's speech tokens may retain stylistic information like timbre due to the absence of supervision during tokenizer training. Second, R-VC’s mask transformer duration model excels at modeling duration, enabling accurate rhythm replication and facilitating emotion transfer, providing more flexible control over rhythm than baseline models which maintain the source speech’s rhythm.



\subsection{Rhythm Control Ability Evaluation}
While the improvement in emotion style transfer demonstrates the effectiveness of our model in rhythm modeling, it remains an indirect measure. Therefore, we introduce an intuitive evaluation metric, \emph{speaking rate}, to quantify rhythm control. Following \cite{ji2024textrolspeech}, we compute speaking rate as phonemes per second (PPS) after applying voice activity detection (VAD) to remove silence. We categorize the speech into three rhythm levels: slow ($<7.98$ PPS), fast ($>14.47$ PPS), and normal ($7.98 \leq x \leq 14.47$ PPS).

We divide all 2,620 test-clean samples from the English subset of the Multilingual LibriSpeech (MLS) dataset into these three categories and construct non-parallel source-target pairs across rhythm levels. For each synthesized utterance, we calculate the PPS, assign it a rhythm label, and evaluate the alignment with the target rhythm.

\begin{table}[h]
\scriptsize
\centering
\begin{tabular}{lcc}
\toprule
\textbf{Target Category} & \textbf{Avg. Speaking Rate (PPS)} & \textbf{Accuracy} \\
\midrule
Slow (151)    & 7.561  & 86.0\% \\
Normal (2071) & 11.903 & 92.3\% \\
Fast (398)    & 14.257 & 80.7\% \\
\midrule
\textbf{Overall} & -      & \textbf{90.2\%} \\
\bottomrule
\end{tabular}
\caption{Speaking rate statistics of generated speech for different target rhythm categories. Numbers in parentheses indicate sample sizes.}
\label{table:speaking_rate}
\end{table}

As shown in Table~\ref{table:speaking_rate}, the average speaking rate of the generated speech closely matches the target category, with a rhythm classification accuracy of 90.2\%. This result demonstrates the strong rhythm control ability of our proposed masked duration model. It is worth noting that baseline models retain the source speech’s duration and thus cannot adapt to the target rhythm. Therefore, we exclude baseline results from this analysis. Audio examples are provided in the “Rhythm Control Demo (Rebuttal)” section on our demo page for qualitative reference.

\begin{table}[th]
\small
\tabcolsep=2.5pt
\centering
\begin{tabular}{l|cccc}
\toprule
\bfseries Method & \bfseries {WER} & \bfseries {SECS} & \bfseries UTMOS & \bfseries EMO \\ 
\midrule
R-VC &\bfseries6.95&\bfseries0.880&3.85&\bfseries0.590\\
\midrule
w/o $dur$ &  7.03&0.878&\bfseries3.86&0.425\\
w/o $spk\_v1$ &  6.99&0.860&3.75&0.571\\
w/o $spk\_v2$ &  8.24&0.873&3.69&0.477\\
w/o $perturb$ &  7.28&0.869&3.78&0.580\\
w/i $sdur$ &  9.86&0.872&3.58&0.528\\
\bottomrule
\end{tabular}
\caption{Ablation Studies on rhythm modeling, timbre condition and data perturbation methods.}
\label{table3}
\vspace{-0.2in}
\end{table}

\subsection{Ablation Studies}
We conducted ablation experiments on the ESD to validate the effectiveness of our system's design and each module, including: removing the duration module ($w/o , dur$), separately excluding global speaker conditioning from the DiT decoder and the duration model ($w/o \, spk\_v1$ and $w/o \, spk\_v2$), omitting pitch perturbation before content token extraction ($w/o \, perturb$). Additionally, following \cite{chen2024f5}, we summed the unit-level duration to obtain sentence-level duration and trained the DiT decoder with a padding filter token for deduplicated content tokens ($w/i , sdur$).

The results are shown in Table \ref{table3}. The ablation study reveals that: 1) Removing the duration module significantly impairs emotion transfer (emotion score -0.165), as the DiT decoder preserves source rather than target speech rhythm, which is crucial for emotional expression. Notably, some inaccurate duration predictions may also reduce the model's naturalness compared to those without the duration module. 2) Using sentence-level duration as a global constraint makes the DiT decoder struggle with learning alignment between text and speech, leading to unstable performance and increased WER. 3) The absence of global speaker conditioning degrades overall performance, particularly affecting timbre similarity in R-VC, leveraging speaker verification system's timbre representation with the target prompt proves beneficial, effectively capturing both time-invariant and time-varying speaker characteristics while strengthening timbre modeling. 4) Removing speaker conditioning from the duration model impacts WER, UTMOS, and emotion scores, since rhythm or duration patterns are inherently tied to both content and speaker characteristics. 5) Data perturbation before token extraction effectively mitigates timbre leakage by filtering content-irrelevant information.

\begin{figure}[]
\centering
\includegraphics[height=3.0cm, width=5.5cm]{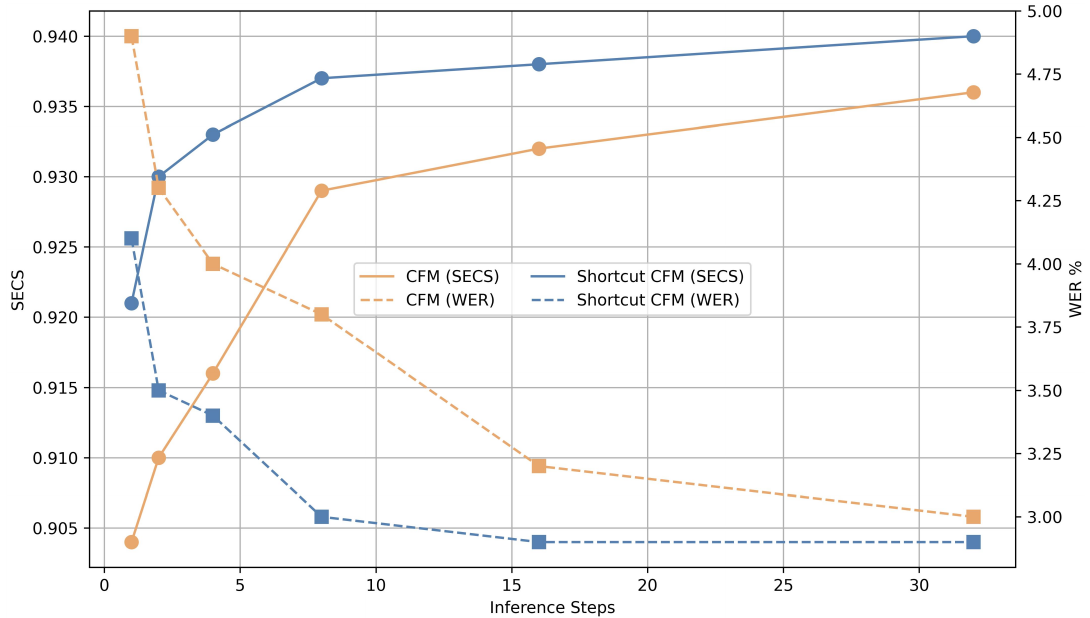}
\caption{Results with different inference steps.}
\label{nfe_wer_sces}
\vspace{-0.2in}
\end{figure}

\textbf{Varying generation steps}
We compared the proposed shortcut flow matching with naive flow matching at different sampling step counts. As shown in Figure \ref{nfe_wer_sces}, although the original CFM performs well with sufficient steps ($\geq$ 10), its performance drops sharply at lower step counts (in terms of WER and SECS). In contrast, our shortcut CFM achieves matching performance with just two steps, maintaining minimal performance loss as step count decreases. This highlights the method's superior efficiency in generating high-quality speech with fewer steps.

\begin{table}[ht]
\centering
\scriptsize
\tabcolsep=2.0pt
\begin{tabular}{l|ccc|ccc}
\toprule
\bfseries \multirow{2}{*}{Decoder(NFE=2)} & \multicolumn{3}{c}{\bfseries LibriTTS} & \multicolumn{3}{c}{\bfseries MLS EN} \\ \cmidrule{2-7} 
& WER & SECS & UTMOS & WER & SECS & UTMOS \\
\midrule
U-Net  & \bfseries4.58&0.892&\bfseries3.99&3.60&0.925 &4.08  \\
DiT  & 4.61&\bfseries0.897&3.97&\bfseries3.51&\bfseries0.930 &\bfseries4.10  \\
\bottomrule
\end{tabular}
\caption{Ablation Studies on different architectures.}
\label{table4}
\vspace{-0.1in}
\end{table}

\textbf{Ablation of Model Architecture}
We evaluated U-Net and Diffusion Transformer decoders with 100MB and 300MB parameters on the LibriTTS and MLS datasets (Table \ref{table4}). While both architectures showed similar performance, the U-Net generated clearer audio with slight improvements in WER and UTMOS at smaller parameter sizes. As the parameter size and dataset scale increased, the DiT outperformed the U-Net across all metrics, particularly in speaker similarity. The analysis of the duration model's performance, ablation of the self-consistency loss hyperparameter $k$, the effect of speech prompt lengths, and UTMOS performance across inference steps are in Appendix \ref{details_ablation}.

\section{Conclusion}
In this paper, we propose R-VC, an efficient zero-shot voice conversion system that enables flexible control over rhythm and speaker identity. By leveraging a Diffusion Transformer with a shortcut flow matching strategy, and a Mask Generative Transformer for in-context duration modeling, R-VC not only adapts the duration of linguistic content to various speaking styles but also captures both time-invariant and time-varying speaker characteristics. This approach enables high-quality speech generation with minimal sampling steps, significantly enhancing both efficiency and performance. 

\section{Acknowledgement}
This work was supported by computational resources and technical assistance provided by Huawei Cloud. We sincerely thank them for their valuable support.

\section{Limitations and Future Work}
In this section, we discuss the limitations and challenges of our R-VC model. In R-VC, we use a mask transformer model for rhythm modeling. However, our experiments revealed that there is a certain probability of inaccurate duration predictions, which leads to unnatural phenomena in the generated speech, such as overextended pronunciations. We believe that this fine-grained duration model poses challenges to the stability of our system. Therefore, we also explored a sentence-level duration strategy in this work, where the model implicitly learns the duration alignment of tokens with the mel-spectrogram. However, this approach yielded suboptimal results and exhibited even worse stability. Moving forward, we aim to explore more robust duration modeling strategies to enhance the style transfer capability of the zero-shot voice conversion model.

\bibliography{custom}

\appendix

\label{sec:appendix}
\section{Data Perturbation}
\label{pitch_data}
We propose to perturb the pitch information in the input waveform $x$ using three functions: 1) formant shifting $fs$, 2) pitch randomization $pr$, and 3) random frequency shaping using a parametric equalizer $peq$. The chain function is employed to randomly shift the pitch value of the original speech $S$ denoted as: $F = \text{fs}(\text{pr}(\text{peq}(S)))$.
The specific methods are as follows:
\begin{itemize}
    \item For \textbf{$fs$}, a formant shifting ratio is sampled uniformly from \(\text{Unif}(1, 1.4)\). After sampling the ratio, we again randomly decide whether to take the reciprocal of the sampled ratio or not.
    \item For \textbf{$pr$}, a pitch shift ratio and pitch range ratio are sampled uniformly from \(\text{Unif}(1, 2)\) and \(\text{Unif}(1, 1.5)\), respectively. Again, we randomly decide whether to take the reciprocal of the sampled ratios or not. For more details on formant shifting and pitch randomization, please refer to Parselmouth \footnote{\url{https://github.com/YannickJadoul/Parselmouth}}.
    \item \textbf{$peq$} represents a serial composition of low-shelving, peaking, and high-shelving filters. We use one low-shelving (HLS), one high-shelving (HHS), and eight peaking filters (HPeak).
\end{itemize}

\label{sec:appendix}
\section{Flow Matching}
\label{cfm_appendix}
\subsection{Conditional Flow Matching}
Continuous Normalizing Flows (CNFs) \cite{chen2018neural} aims to estimate the unknown distribution $q(x)$ of data $x \in \mathbb{R}^d$ by learning the probability path from a simple prior distribution $p_0$ to a data distribution $p_1 \approx q$. This mapping can be further taken as a time-dependent changing process of probability density (a.k.a. flow), determined by the ODE:
\begin{equation}
\frac{d}{dt}{\phi_t(x)} = v_t(\phi_t(x)); \quad \phi_0(x) = x.
\end{equation}
where $v_t : [0, 1] \times \mathbb{R}^d \rightarrow \mathbb{R}^d$ is a vector field that generates the flow $\phi_t : [0, 1] \times \mathbb{R}^d \rightarrow \mathbb{R}^d$. We can sample from the approximated data distribution $p_1$ by solving the initial value problem in Eq. (1). Suppose there exists a known vector field $u_t$ that generates a probability path $p_t$ from $p_0$ to $p_1$. The flow matching loss is defined as:
\begin{equation}
L_{FM}(\theta) = E_{t,p_t(x)}\|u_t(x) - v_t(x; \theta)\|^2.
\end{equation}
where $t \sim U[0, 1]$ and $v_t(x; \theta)$ is a neural network with parameters $\theta$. However, $L_{FM}$ is uncomputable for lack of prior knowledge of $p_t$ or $v_t$. Luckily, \cite{lipman2022flow} proposed Conditional Flow Matching (CFM) objective presented as: 
\begin{equation}
L_{CFM}(\theta) = E_{t,q(x_1),p_t(x|x_1)}\|u_t(x|x_1) - v_t(x; \theta)\|^2
\end{equation}
By conditioning $p_t$ and $v_t$ on real data $x_1$, they proved that FM and CFM have identical gradients with respect to $\theta$ for training generative model.

In R-VC, an optimal-transport conditional flow matching model (OT-CFM) \cite{lipman2022flow} is employed to learn the distribution of Mel spectrogram and generate samples from gaussian noise. The OT-CFM loss function can be written as:
\begin{equation}
\begin{split}
L(\theta) &= E_{t,q(x_1),p_0(x_0)}\|u^{OT}_t(\phi^{OT}_t(x)|x_1) \\ 
&- v_t(\phi^{OT}_t(x)|\theta)\|^2
\end{split}
\end{equation}
by defining $\phi^{OT}_t(x) = (1 - (1 - \sigma_{\text{min}})t)x_0 + tx_1$ as the flow from $x_0$ to $x_1$ where each datum $x_1$ is matched to a random sample $x_0 \sim N(0, I)$. 


\subsection{Shortcut Flow Matching}
Flow-matching trains an ODE to map noise to data along curved trajectories. However, taking large sampling steps naively results in significant discretization errors and, in single-step scenarios, can lead to catastrophic failures. By conditioning on $d$, shortcut flow matching models can anticipate future curvature and accurately jump to the correct next point, avoiding divergence. We define the normalized direction from $x_t$ to the next target point $x'_{t+d}$ as the shortcut $s(x_t, t, d)$:
\begin{equation}
x^{'}_{t+d} = x_t + s(x_t, t, d) \cdot d
\end{equation}
The objective is to train a shortcut model $s_\theta(x_t, t, d)$ to learn shortcuts for all combinations of $x_t$, $t$, and $d$. Shortcut models generalize flow-matching models to larger steps by learning to handle transitions beyond instantaneous velocity. As $d \to 0$, the shortcut aligns with the flow.

A direct approach to compute training targets involves simulating the ODE with small step sizes \cite{luhman2021knowledge}, but this is computationally expensive. An alternative exploits the self-consistency of shortcut models: a single shortcut step is equivalent to two consecutive steps of half the size.
\begin{equation}
s(x_t, t, 2d) = \frac{1}{2} \left( s(x_t, t, d) + s(x'_{t+d}, t, d) \right)
\end{equation}
Shortcut models can be trained using self-consistency targets for $d > 0$ and the flow-matching loss for $d = 0$. The model can, in principle, be trained on any distribution $d \sim p(d)$. In practice, the batch is split: one fraction uses $d = 0$, while the other uses randomly sampled $d > 0$. This leads to the combined shortcut model loss function.
\begin{equation}
\begin{split}
L_{\small \text{S-CFM}}(\theta) = \mathbb{E}_{p_0(x_0), p_1(x_1), (t,d)} [ \| s_\theta(x_t, t, 0) -\\ (x_1 - x_0) \|^2 \, + \| s_\theta(x_t, t, 2d) - s_{\text{target}} \|^2 ] \nonumber
\end{split}
\end{equation}
where $s_{\text{target}} = \frac{1}{2}s_\theta(x_t, t, d) + \frac{1}{2}s_\theta(x'_{t+d}, t, d)$, where $x'_{t+d} = x_t + s_\theta(x_t, t, d)d$. This objective intuitively trains the model to map noise to data in a way that remains consistent across any sequence of step sizes, including a single large step. This training objective combines a flow-matching objective and a self-consistency objective, which are jointly optimized during training. To enhance training efficiency, we construct a batch by mixing $k$ flow-matching targets with $(1-k)$ self-consistency targets.

\label{sec:appendix}
\section{Details of Models}
\label{details_model}
The ground truth mel-spectrograms are generated from the raw waveform with a frame size of 1024, a hop size of 256, and 80 channels at a sampling rate of 22.05 kHz. An open-source HiFiGAN vocoder \footnote{\url{https://www.modelscope.cn/models/iic/CosyVoice-300M}} is used to synthesize speech from the mel-spectrograms. The Diffusion Transformer Decoder consists of a 22-layer Transformer architecture with 16 attention heads and 1,024 embedding dimensions, totaling 300M parameters. The Mask Transformer Duration Model follows a similar parameter configuration, but features a hidden dimension of 512 and comprises a total of 62M parameters. A random 70\% to 100\% of mel frames is masked for infilling task training. The classifier-free guidance scale $ \alpha $ is set to 0.2 during training and 0.7 during inference, as referred to in \cite{du2024cosyvoice}. The shortcut flow matching decoder consists of two training objectives. In each training batch, 30\% of the elements are used to construct self-consistency targets, while the remaining 70\% are used for flow matching targets, ensuring an efficient trade-off between training efficiency and model performance. We use 128 discrete time steps to approximate the ODE, corresponding to 8 potential shortcut paths at each training step.

\label{sec:appendix}
\section{Detailed Experiment Settings}
\label{details_exp}
\subsection{Details in Subjective Evaluation}
We randomly select 50 sentences from the test set and perform the subjective evaluation on Amazon Mechanical Turk (MTurk). Each generated audio has been listened to by at least 10 native listeners. For QMOS evaluations, the listeners are instructed to focus on assessing the audio quality and naturalness while disregarding any differences in styles (such as timbre, emotion, and pronunciation). Conversely, for SMOS evaluations, the listeners are instructed to concentrate on evaluating the speaker similarity to the audio prompt, while disregarding differences in content or audio quality. For the QMOS, SMOS evaluations, each listener is asked to rate different speech samples using a Likert scale ranging from 1 to 5. 

\subsection{Details in Objective Evaluation}
To measure the speaker similarity (SECS), we use the WavLM \cite{chen2022wavlm} model fine-tuned for speaker verification from \footnote{\url{https://huggingface.co/microsoft/ wavlm-base-plus-sv}} to extract the speaker embedding. Then the cosine similarity between the synthesized speech’s speaker embedding and the prompt speech’s speaker embedding is calculated as the speaker similarity score. For word error rate (WER) and character error rate (CER) metrics, we use the Whisper model \footnote{\url{https://huggingface.co/openai/whisper-large-v3}}. 

\begin{table}[th]
\scriptsize
\tabcolsep=2.5pt
\centering
\begin{tabular}{l|ccc}
\toprule
\bfseries Method & \bfseries {ACC\_Absolute} & \bfseries {ACC\_1} & \bfseries {ACC\_5} \\ 
\midrule
masked druation model (w/i prompt) &\bfseries0.744&\bfseries0.986&\bfseries1.0\\
masked druation model (w/o prompt) &0.723&0.978&1.0\\
conformer duration model (w/i prompt) &0.679&0.952&0.996\\
conformer duration model (w/o prompt) &0.665&0.939&0.993\\
\bottomrule
\end{tabular}
\caption{The performance of the masked duration model.}
\label{table5}
\vspace{-0.1in}
\end{table}

\label{sec:appendix}
\section{Additional Experiments}
\label{details_ablation}
\textbf{Masked Duration Model Performance} In fact, the evaluation of the duration model is closely tied to the Diffusion Transformer, as the rhythm control exhibited by the generated speech directly reflects the effectiveness of the masked duration model. Here, we further evaluate the duration model's prediction accuracy to provide a more comprehensive view of its performance. The experimental results are shown in Table \ref{table5}. The Conformer duration model refers to a non-autoregressive Conformer for duration modeling. $ACC\_Absolute$ measures the prediction accuracy of each de-duplicated HuBERT token's duration, while $ACC\_1$ and $ACC\_5$ allow prediction errors within one token (0.02s) and five tokens (0.1s), respectively. $W/i\ prompt$ and $w/o\ prompt$ indicate whether prompt conditions (duration and speaker) are used during prediction. As shown in the table, on the test-clean dataset with 2,620 unseen samples, our duration model outperforms the baseline in all three metrics. Specifically, it achieves 98.6\% accuracy within a 0.02s margin and 100\% accuracy within a 0.1s margin, demonstrating the effectiveness of the masked generative duration model. 

\textbf{Evaluation on Seed-TTS} We further evaluate zero-shot voice conversion on the English subset of the Seed-TTS test set \footnote{\url{https://github.com/BytedanceSpeech/seed-tts-eval}} (samples from the Common Voice dataset, containing more complex audio types). The results are summarized in Table~\ref{table:seedtts_results}. The proposed model achieves comparable timbre similarity to Vanilla CFM and CosyVoice-VC with fewer sampling steps, and attains lower WER while significantly reducing latency.

\begin{table}[h]
\small
\centering
\begin{tabular}{lcc}
\toprule
Model & SECS & WER \\
\midrule
DiffHier            & 0.8531 & 0.0589 \\
CosyVoice-VC        & 0.9129 & 0.0610 \\
FAcodec             & 0.8890 & 0.0569 \\
HierSpeech++        & 0.8670 & 0.0451 \\
SEF-VC              & 0.8661 & 0.0494 \\
CFM (NFE=10)        & 0.9110 & 0.0407 \\
Shortcut CFM (NFE=2)& 0.9102 & 0.0418 \\
\bottomrule
\end{tabular}
\caption{Zero-shot voice conversion performance on the Seed-TTS English subset.}
\label{table:seedtts_results}
\vspace{-0.1in}
\end{table}

\begin{table}[th]
\small
\centering
\begin{tabular}{l|ccc}
\toprule
\bfseries 1-k & \bfseries {WER} & \bfseries {SECS} & \bfseries {UTMOS} \\ 
\midrule
10\% &3.88&0.915&3.92\\
20\% &3.62&0.923&4.05\\
30\% &\bfseries3.51&0.930&\bfseries4.10\\
40\% &3.54&\bfseries0.933&4.09\\
\bottomrule
\end{tabular}
\caption{Results of different hyperparameter $k$.}
\label{table6}
\vspace{-0.1in}
\end{table}

\textbf{Additional Ablations} We conducted an ablation analysis on the hyperparameter $k$ under the NFE=2 setting. During each training step, we selected $(1-k)$ proportion of samples to construct the self-consistency training objective and compute the loss. As demonstrated in Table \ref{table6}, a smaller proportion of self-consistency training objective leads to insufficient training of shortcut flow matching, manifesting in suboptimal generation results with fewer sampling steps similar to vanilla CFM. While increasing this proportion effectively enhances the model's performance with fewer steps, it simultaneously introduces additional computational overhead, as the construction of self-consistency training objective necessitates two forward passes. Balancing training efficiency and performance considerations, we ultimately determined that setting 30\% represents an optimal parameter choice.

\begin{figure}[]
\centering
\includegraphics[height=4.0cm, width=6cm]{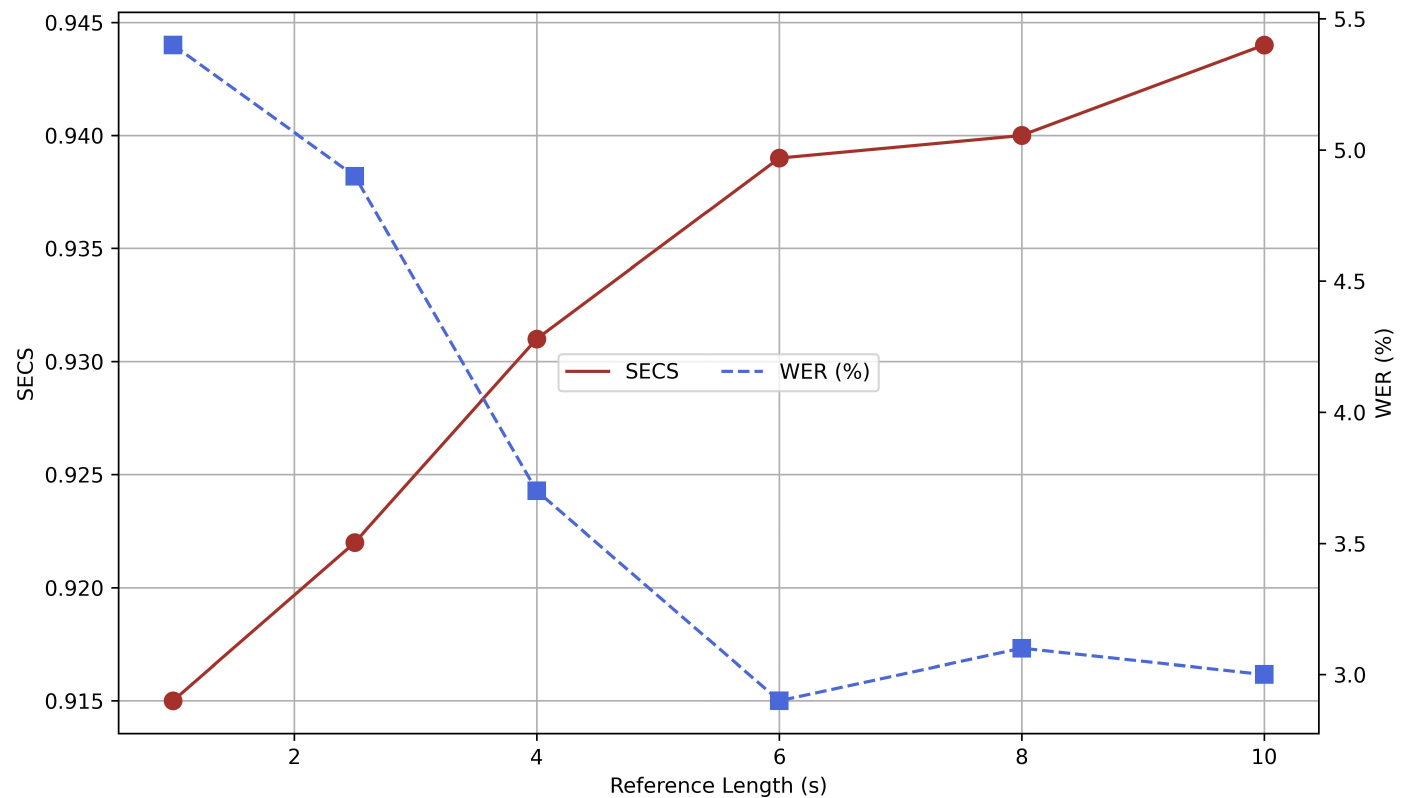}
\caption{The influence of references of different lengths on zero-shot VC.}
\label{ref_prompt}
\vspace{-0.1in}
\end{figure}

\begin{figure}[]
\centering
\includegraphics[height=4.0cm, width=6cm]{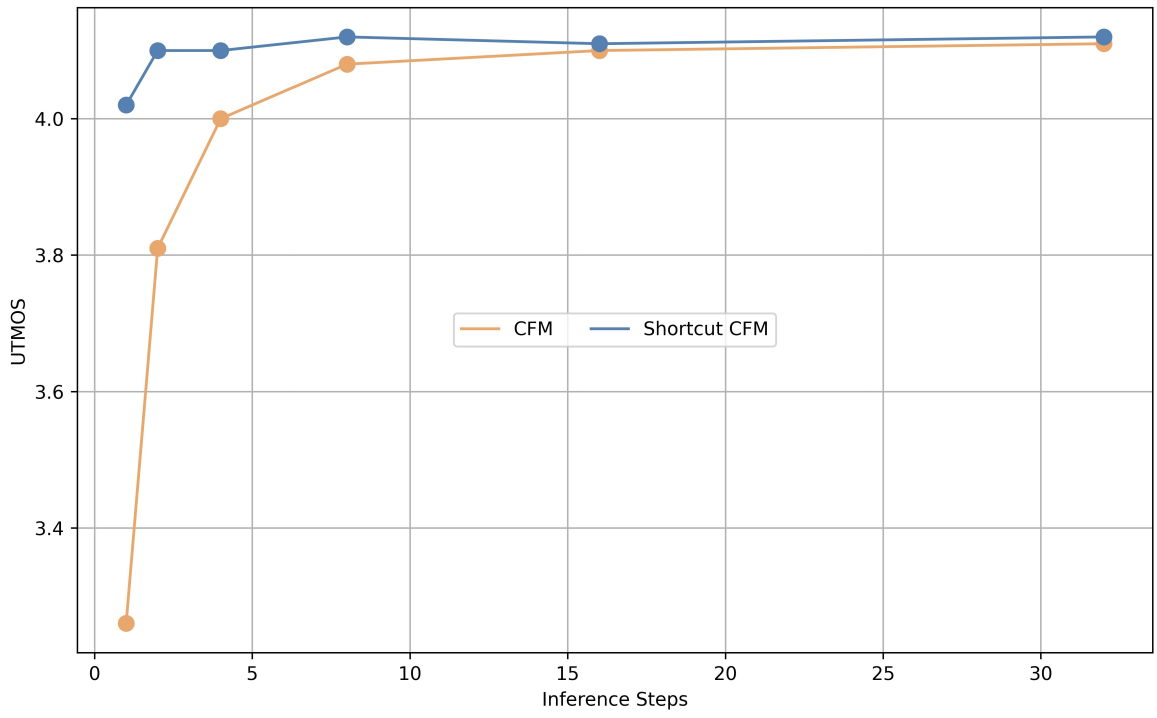}
\caption{The UTMOS performance under different inference steps.}
\label{nfe_utmos}
\vspace{-0.2in}
\end{figure}

We also present the VC performance of our model on the test-clean dataset using speaker prompts of varying lengths. As shown in Figure \ref{ref_prompt} and Table \ref{table1}, even with prompts shorter than 2 seconds, our model achieves higher speaker similarity compared to most baseline models. Speaker similarity improves significantly with longer reference speaker prompts before leveling off. A similar trend is observed for WER, indicating that longer prompts enhance the intelligibility of the generated speech. Moreover, we demonstrate the performance of our proposed method and the vanilla CFM method in terms of UTMOS across varying sampling steps. As shown in Figure \ref{nfe_utmos}, our method significantly outperforms the baseline flow matching methods in terms of audio perceptual quality with fewer sampling steps, indicating the model's ability to generate high-quality speech with minimal sampling steps.



\end{document}